\documentclass[preprint,showpacs,preprintnumbers,amsmath,amssymb]{revtex4}

\usepackage{graphicx}
\usepackage{dcolumn}
\usepackage{bm}

\begin{document}

\preprint{APS/123-QED}

\title{Probing the shape of atoms in real space}

\author{M. Herz}
\author{F. J. Giessibl}
\email{Franz.Giessibl@physik.uni-augsburg.de}
\author{J. Mannhart}
\homepage{http://www.Physik.Uni-Augsburg.DE/exp6}

\affiliation{%
Universit\"at Augsburg, Institute of Physics, Electronic
Correlations and Magnetism, Experimentalphysik VI,
Universit\"atsstrasse 1, D-86135 Augsburg, Germany.
}%

\date{submitted to PRB March 7 2003, revised version April 23 2003}

\begin{abstract}
The structure of single atoms in real space is investigated by
scanning tunneling microscopy. Very high resolution is possible by
a dramatic reduction of the tip-sample distance. The instabilities
which are normally encountered when using small tip-sample
distances are avoided by oscillating the tip of the scanning
tunneling microscope vertically with respect to the sample. The
surface atoms of Si(111)-(7$\times$7) with their well-known
electronic configuration are used to image individual samarium,
cobalt, iron and silicon atoms. The resulting images resemble the
charge density corresponding to 4f, 3d and 3p atomic orbitals.
\end{abstract}

\maketitle

\begin{figure}
\includegraphics[width=8.6cm]{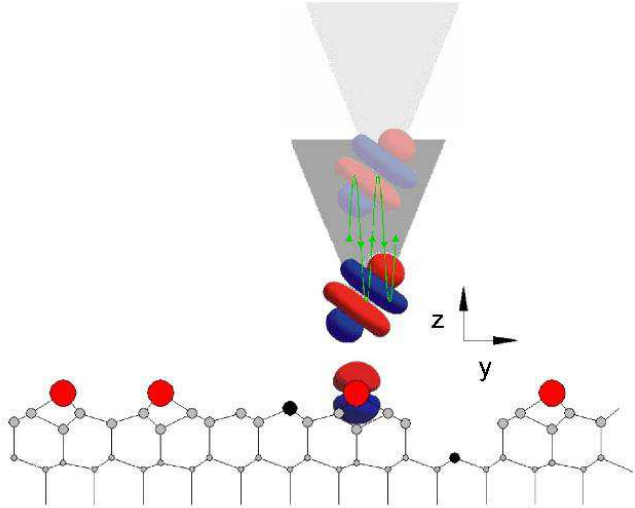}
\caption{\label{fig:Fig1} (Color online) Principle of dynamic STM.
The tip is mounted on a quartz cantilever (not shown here, see
\cite{Giessibl1998} for details) with a stiffness of 1800 N/m
which oscillates at a fixed amplitude $A \approx 0.3$ nm. The
unperturbed resonance frequency is $f_0 \approx 20$ kHz.
Conservative components in the tip-sample force are measured by
the change in the cantilevers eigenfrequency they are causing
\cite{Giessibl2000}. Non-conservative components in the tip-sample
force are measured by monitoring the dissipation energy which has
to be supplied to the cantilever to maintain a constant amplitude
\cite{Giessibl2002}.}
\end{figure}

Democritos of Abdera (460 BC - 370 BC) predicted the existence of
atoms and speculated that atoms come in different sizes and shapes
\cite{Heisenberg}. Today, we know that the size and shape of an
atom depends on its chemical species and the way it is bonded to
neighboring atoms. Atoms consist of the tiny, positively charged
nucleus and electrons with negative charge $-e$ which orbit the
nucleus with a speed of a few percent of the speed of light.
Because of the uncertainty principle of quantum mechanics,
articulated by Werner Heisenberg in 1927, we can not observe
single electrons within their orbit. However, quantum mechanics
specifies the probability of finding an electron at position
$\textbf{x}$ relative to the nucleus. This probability is
determined by $|\psi(\textbf{x})|^2$, where $\psi(\textbf{x})$ is
the wave function of the electron given by Schr\"{o}dinger's
equation \cite{Dirac}. The product of $-e$ and
$|\psi(\textbf{x})|^2$  is usually interpreted as a charge
density, because the electrons in an atom move so fast that the
forces they exert to other charges are essentially equal to the
forces caused by a static charge distribution
$-e|\psi(\textbf{x})|^2$. For the electrons of an atom, the
solutions to Schr\"{o}dinger's equation are wavefunctions
$\psi_{nlm\sigma}(\textbf{x})$ defined by four quantum numbers:
principal quantum number $n$, angular momentum $l$, $z$-component
of angular momentum $m$ and spin $\sigma$ \cite{Baym}.

The shape of atoms, as given by their total charge density, can be
determined by X-ray- or electron-scattering if they are arranged
periodically in a crystal. X-ray scattering allows to measure the
roughly spherical charge density of ions in alkali-halides
\cite{Schoknecht} or the non-spherical structure of atoms
crystallized in a diamond lattice \cite{Yang}. Recently, d-orbital
holes were found in the charge density of copper-oxide compounds
by transmission electron microscopy measurements \cite{Zuo}.
Observing the charge density of single atoms in real space became
feasible in 1982 with the invention of scanning tunneling
microscopy (STM) \cite{Binnig}. In contrast to the techniques
listed above, STM does not probe the total charge density of an
atom, but the charge density at the Fermi level $\rho_{Fermi}$
\cite{Tersoff,Chen}. For negative sample bias voltage, the charge
density of the energetically highest occupied electron states is
imaged, while for positive sample bias, the charge density of the
lowest unoccupied states is imaged. If the states at the Fermi
level have an s-type symmetry, the constant-density surface
$\rho_{Fermi}(\textbf{x}) = c$ is composed of spheres centered at
the nuclei of the surface atoms. If $\rho_{Fermi}$ is built from
atomic states with higher angular momentum, $\rho_{Fermi}$ can
have a more complex shape, in particular for large values of
$\rho_{Fermi}$ in close proximity to the atoms. Thus, if higher
momentum states are present at the Fermi energy, the corresponding
STM images are expected to reflect their shape, such that the
shapes of atomic images are non-spherical and display orbital
substructures. However, it is important to note that orbitals in
the sense of mathematical constructs with an amplitude and phase
can not be observed, but the density of charges within a certain
energy range corresponding to specific orbitals, see
\cite{Tersoff,Chen} and the discussion \cite{Wang2000} and
references therein.

In spite of the large amount of standard STM data gathered, the
observation of atomic substructures linked to atomic
higher-momentum states has not been reported, instead STM images
of atoms always appear more or less parabolic. Interestingly,
tunneling channels originating from higher orbital momentum states
have been found in tunneling experiments involving mechanically
controllable break junction techniques \cite{Scheer1998}, where
tunneling gap widths of the order of bulk next-neighbor distances
can be realized. However, imaging is not possible with the break
junction technique. While {\lq subatomic\rq} resolution, showing
two maxima in the image of a single atom linked to sp$^3$-type
orbitals has been observed by force microscopy
\cite{Giessibl2000Science, Giessibl2001a}, features within single
atoms have not yet been observed by STM. After a presentation of
the work reported in \cite{Giessibl2000Science} by a member of our
team, G. Doyen from the University of Munich insisted that orbital
structures should be visible in STM, if these structures can be
imaged by force microscopy. Also, C.J. Chen (chapter 7 in
\cite{Chen}) discusses the similarity of the functional dependence
of tunneling current and attractive part of the tip-sample force.
These theoretical arguments and the experimental appearance of
higher momentum states in AFM- and break-junction-tunneling-
experiments inspired us to think about possible reasons for the
absence of subatomic resolution in standard STM
\cite{Giessibl2001a}.

One likely cause is the large tip-sample separation in standard
STM (0.5 - 0.8 nm). Even if the Fermi level in a sample or tip was
occupied with higher momentum states, the contour lines of the
electron density show at this large distance only one single
maximum per surface- and tip-atom. A second possible cause is the
geometry of the electronic states participating in the tunneling
process. The valence bands of solids can often be constructed from
atomic states by Linear Combination of Atomic Orbitals (LCAO). As
a consequence, spatial features are expected to be present in
individual tunneling channels \cite{Chen}. Because of various
effects \cite{Baym}, the energy of atomic (and bulk-) states
depends on the angular momentum $l$. Thus, if large tunneling
voltages are used, many states can contribute to the tunneling
current, smearing out spatial features. We concluded that small
bias voltages and very small tip-sample distances are required for
enhanced resolution.

In standard STM, the tip sample distance decreases when the
tunneling impedance, given by the bias voltage divided by the
setpoint of the tunneling current is decreased. In STM operation,
the setpoint of the tunneling impedance has to satisfy a
compromise. If the setpoint of the tunneling current is too small,
the corrugation in the image is small and the noise level is high.
If the current setpoint is too high (and/or bias voltage too low),
the tip-sample forces become excessive and destruction of tip and
sample occurs when scanning at high loading forces. However, when
the lateral scan is stopped to allow for current-versus-distance
spectroscopy experiments, small tip-sample distances can be
realized without tip or surface destruction. A recent theoretical
study \cite{Jarvis2001} has shown that single atom tips can
withstand repulsive forces up to 30 nN without damage. However,
pushing the tip laterally over a surface at high loading forces
damages both tip and sample. A stop-and-go approach where the scan
is stopped, the tip is lowered until it reaches its high current
setpoint, retracted and moved laterally to start over again would
remedy this situation. We found experimentally that a much simpler
and faster approach is possible by deliberately vibrating the tip
vertically (see Fig.~\ref{fig:Fig1}). Because of the strong
distance dependence of the tunneling current, the current only
flows when the tip is close to its lower turnaround point. The
lateral motion of the tip during this time when it is very close
to the sample is given approximately by the scanning speed times
less than half of the oscillation period, amounting to only a few
pm.

A similar concept already proved successful in tapping force
microscopy, where the oscillation of the tip also reduces lateral
forces and decreases tip wear \cite{Zhong1993}. The tip
oscillation in \textit{dynamic STM} reduces or even prevents the
wear which normally occurs when scanning a probe tip across a
surface with repulsive load forces.

In our experiments, the vibration of the tip is done by mounting
it on a quartz force sensor which oscillates at frequency $f_0
\approx$ 20 kHz, as described in detail in \cite{Giessibl1998}.
While the oscillation could also be fashioned with a standard STM,
mounting the tip on a cantilever allows to monitor two more
important observables: the conservative and the dissipative
components of the tip-sample force. The forces which occur during
tunneling have been measured before with a different technique,
albeit not with atomic resolution \cite{Durig}. Measuring the
conservative force component yields information about the
tip-sample distance. Monitoring the dissipative component allows
to assess if tip and sample atoms are deflected strongly from
their equilibrium positions \cite{Giessibl2002}. The STM images
are recorded in the topographic mode, where the average tunneling
current (averaged over a time of approx. 30 oscillation cycles) is
kept constant by adjusting the $z$-position of the tip
accordingly. For an exponentially decaying tunneling current $I(z)
= I_0 exp(-2{\kappa}z)$, the average tunneling current $I_{av}$ is
smaller than the peak tunneling current $I_{pk}$ with
$I_{av}/I_{pk} \approx (2\pi\kappa A)^{-0.5}$ \cite{Giessibl2000}.
In this publication, we present measurements performed by dynamic
STM and report the observation of  atomic images containing
symmetry features linked to p, d and f orbitals of Si, Co and Sm
atoms. All experiments have been performed in ultrahigh vacuum at
a pressure of $p\approx10^{-8}$ Pa and ambient temperature
$T\approx$ 300 K.

In STM, the role of tip and sample is perfectly interchangeable -
the image is a convolution of tip and sample states. For zero
temperature, the tunneling conductivity depends on the matrix
elements $M_{\mu\nu}$ in a first order calculation as

\begin{eqnarray}
\frac{I_t}{V}=(2\pi)^2 G_0 \sum_{\mu\nu}^{}D_{S\mu} (E_F) D_{T\nu}
(E_F) \vert M_{\mu\nu}\vert^2 \label{eq:quant}
\end{eqnarray}

with the conductance quantum $G_0=2e^2/h$. Here, the densities of
states $D_{S\mu}$ and $D_{T\nu}$ of the corresponding band states
of sample and tip as well as the matrix elements $M_{\mu\nu}$ are
assumed to be constant in the energy range $eV$ around the Fermi
level $E_F$.

\begin{figure}
\includegraphics[width=16cm]{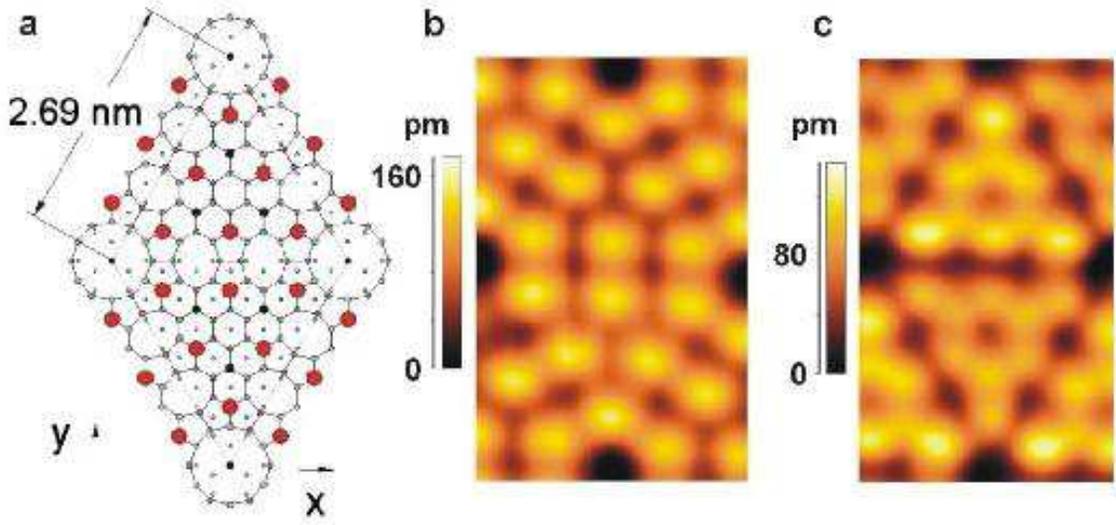}
\caption{\label{fig:Fig2} (Color online) (a) Surface unit cell of
Si (111)-(7x7) after the dimer-adatom-stacking fault model. The
diamond shaped unit cell has characteristic features: deep
cornerholes at the corners of the diamond (see Fig. 1 for a cross
section of the corner hole along the long diagonal of the unit
cell), 18 dimer atoms, 12 adatoms and 6 restatoms. One triangular
halve of the unit cell develops a stacking fault. The surface
atoms (adatoms) are printed in red. Each adatom is bonded to the
layer underneath by three covalent bonds. The forth valence
electron forms a dangling bond (3sp$^3$ state), which sticks out
perpendicular to the surface. The rest atoms, printed in full
black circles, are located more than 100 pm below the adatoms. (b)
Standard STM image of the empty states on the Si
(111)-(7$\times$7) surface (sample bias 2.4 V, tunneling current 1
nA). (c) Standard STM image of the filled states on the Si
(111)-(7$\times$7) surface (sample bias -2.4V, tunneling current 1
nA). In the filled state image, the upper half of the
diamond-shaped unit cell appears to be slightly higher than the
lower half, caused by the stacking fault. In images b and c, the
image of a single adatom is a paraboloid with a radius of
$\approx$ 0.5 nm. Ir tips have been used for both images.}
\end{figure}

In Bardeen's treatment of a tunneling junction \cite{Chen}, the
matrix elements
\begin{math}M_{\mu\nu}\end{math} are evaluated to be

\begin{equation}
M_{\mu\nu}=-\frac{\hbar^2}{2m_e}\int_{\Omega_T}^{}
\left(\chi_\nu^\star\nabla^2\psi_\mu-\psi_\mu\nabla^2\chi_\nu^\star\right)dV
\end{equation}

with the electron's mass $m_e$ and tip and sample states
\begin{math}\chi_\nu\end{math} and \begin{math}\psi_\mu\end{math}, respectively.
The integration extends over the tip volume
\begin{math}\Omega_T\end{math}, see e.g. chapter 2 in \cite{Chen}.
The tunneling process is considered to be spin-independent for the
case of a non-magnetic tip and/or sample. The modulus of the
matrix element does not change if tip and sample states switch
roles, giving rise to the reciprocity principle: \textit{If the
electronic states of the tip and the sample under observation are
interchanged, the image should be the same} - see p. 88 in
\cite{Chen}. Thus a STM image can either be interpreted as one
image of $N$ sample atoms probed by the front atom of the tip or
as $N$ images of a front atom probed by $N$ sample atoms. In this
study, we are interested in the image of the front atom of the
tip. Therefore, the electronic state of the sample atoms needs to
be known. We have chosen a Si (111)-(7$\times$7) surface as sample
(see Fig.~\ref{fig:Fig2} (a), because the electronic states of the
surface atoms (adatoms) in Si (111)-(7$\times$7) are precisely
characterized, so that the tip states can be inferred from the
images.

Each adatom of the Si (111)-(7$\times$7) surface exposes one
sp$^3$-dangling bond oriented perpendicular to the surface. The
large distance between adjacent adatoms (at least 0.67 nm) allows
to clearly separate individual tip atom images. The shape and apex
radius of the images is used to characterize the electronic states
of the tip. The advantage of scanning an area containing many
sample atoms is that potential errors due to multi-tip effects are
easily detected. The observation of single defects on the Si
surface rules out double- or multi-tip effects, and the
observation of deep corner holes offers a proof that the tip is
pointed and sharp. This is not a restriction of our method,
because in principle, well defined tips can also be created
\cite{Giessibl2001a} and unknown sample states could be studied
with these tips. As a reference, standard STM images of Si
(111)-(7$\times$7) are shown in Fig.~\ref{fig:Fig2}.
Fig.~\ref{fig:Fig2} (a) shows the structure of the Si
(111)-(7$\times$7) surface and Figs.~\ref{fig:Fig2} (b), (c) are
STM images taken with an iridium tip, using a tunneling current of
1 nA and a sample bias of +2.4 V (empty states) in Fig.
~\ref{fig:Fig2}b, and a sample bias of -2.4 V (filled states) in
Fig.~\ref{fig:Fig2} (c). The apex of the adatom image can always
be approximated as a paraboloid. The topographic data around the
peak is approximated with $z(x,y) = z_0 - (x-x_{Peak})^2/2R_x -
(y-y_{Peak})^2/2R_y$, thus $R$ is easily determined from the
topographic data. Application of this formula to the measured
profiles in Fig.~\ref{fig:Fig2}b,c yields apex radii of 0.5 nm.

\begin{figure}
\includegraphics[width=16cm]{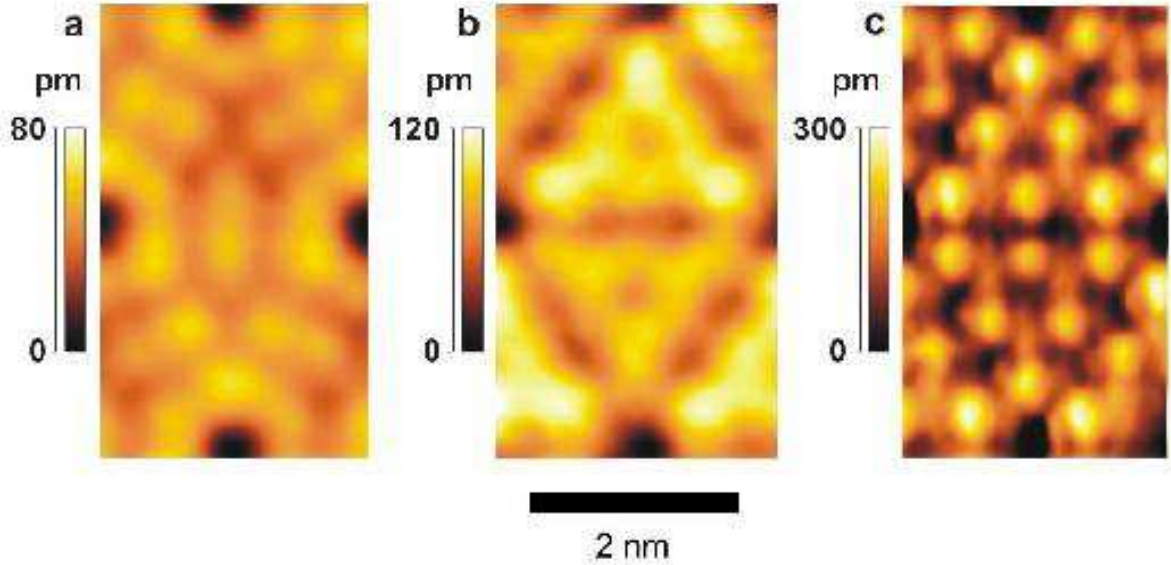}
\caption{\label{fig:Fig3} (Color online) (a) STM images of a Si
(111)-(7$\times$7) surface taken with a Sm tip. A Standard STM,
sample bias: 2 V, $I_t =$ 100 pA, (b) Standard STM, sample bias:
-2 V, $I_t =$ 200 pA, (c) Dynamic STM, sample bias: -1 V, $I_{av}
=$ 100 pA, $A =$ 0.8 nm.}
\end{figure}

Figure~\ref{fig:Fig3} shows the difference between imaging in
standard STM and dynamic STM. A samarium tip was used to image Si
in standard STM in Fig.~\ref{fig:Fig3} (a) and Fig.~\ref{fig:Fig3}
(b), and dynamic STM in Fig.~\ref{fig:Fig3} (c). Because Sm is a
relatively soft material, the tip can only withstand a small
tunneling current of 100 pA at a sample bias of ± 2V in standard
STM. In dynamic STM with $A =$ 0.8 nm, the average current can be
set to 100 pA (yielding a peak current $I_{Pk} \approx 700$ pA) at
a bias of -1 V. Thus, dynamic STM allows to get much closer to the
surface without tip degradation, resulting in improved spatial
resolution. The curvature of an STM image of a single atom is a
characteristic feature. According to a calculation by Chen
\cite{Chen}, the radii of the atom images is a function of the
distance $z$ between tip and sample atoms and the type of atomic
tip and sample states:

\begin{equation}
s-s:R=z\label{eq:rz}
\end{equation}

\begin{equation}
s-p_z:R=\frac{z}{1+1/{\kappa}z}\label{eq:rp}
\end{equation}

\begin{equation}
p_z-d_{z^2}:R=\frac{z}{1+4/{\kappa}z}
\end{equation}

\begin{equation}
p_z-f_{z^3}:R=\frac{z}{1+7/{\kappa}z}\label{eq:r7}
\end{equation}

where $\kappa$ is the decay constant of the tunneling current with
a typical value of $\kappa =$ 1 \AA$^{-1}$.

\begin{figure}
\includegraphics[width=16cm]{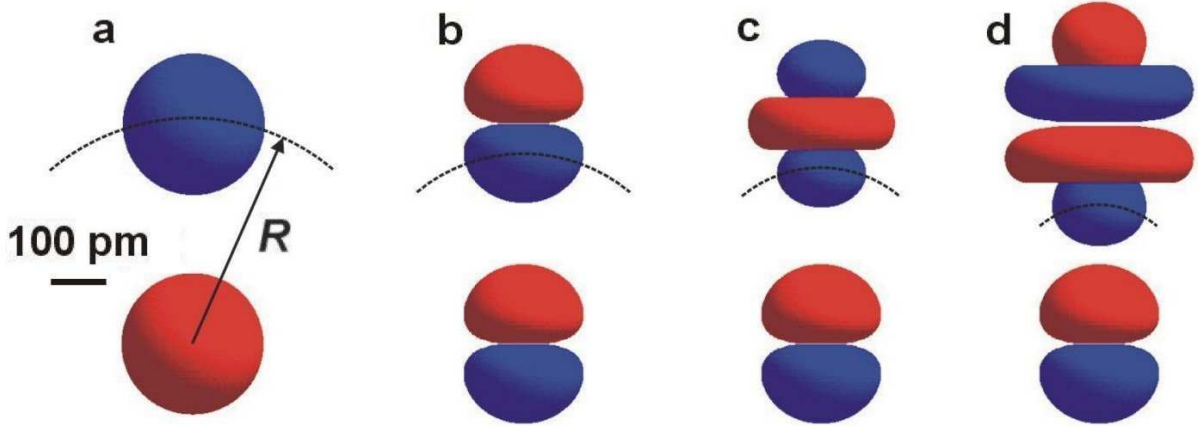}
\caption{\label{fig:Fig4} (Color online) (a) Apparent radius $R$
of a surface atom in standard STM mode as a function of tip and
sample states for a tip-sample distance of $z =$ 0.4 nm.
Calculated with Eqs.~(\ref{eq:rz}-\ref{eq:r7}) after section 6.2
in \cite{Chen}. The tunneling current is assumed to decay with
distance $d$ as $exp(-2\kappa d)$ with $\kappa = 10$ nm$^{-1}$.
{\bf a} Tip state: s orbital, sample state: s orbital. The
apparent radius of the adatom image is given by the distance, thus
$R_a =$ 0.4 nm. (b) Tip state: p$_z$ orbital, sample state: p$_z$
orbital, $R_b =$ 0.27 nm. (c) Tip state: d$_{z^2}$ orbital, sample
state: p$_z$ orbital: $R_c$ = 0.2 nm. (d) Tip state: f$_{z^3}$
orbital, sample state: p$_z$ orbital, R$_d =$ 0.15 nm.}
\end{figure}

Fig.~\ref{fig:Fig4} shows four combinations of tip and sample
atomic states. If tip and sample states are s-type, the radius of
the atom image is simply the distance between tip- and sample
atom. The higher the angular momentum, the smaller the radius of
the atom images. Sharply peaked atom images are observed with tip
states with a large angular momentum. It is evident from
Fig.~\ref{fig:Fig4} that the angular orientation of tip and sample
orbitals is important. Fig.~\ref{fig:Fig4} displays only tip
orbitals oriented in $z$-direction, while the two other p-orbital
orientations p$_x$, p$_y$, the four other d-orbital orientations
d$_{xy}$, d$_{xz}$, d$_{yz}$, d$_{x^2-y^2}$ and the six other
f-orbital orientations can also contribute to the tunneling
current. However, it can be shown that the matrix element between
a p$_z$-sample and p$_z$, d$_{z^2}$ and f$_{z3}$ tip is much
greater than the matrix element between a p$_z$-sample and the
orbitals that are not aligned in $z$-direction.

\begin{figure}
\includegraphics[width=8.6cm]{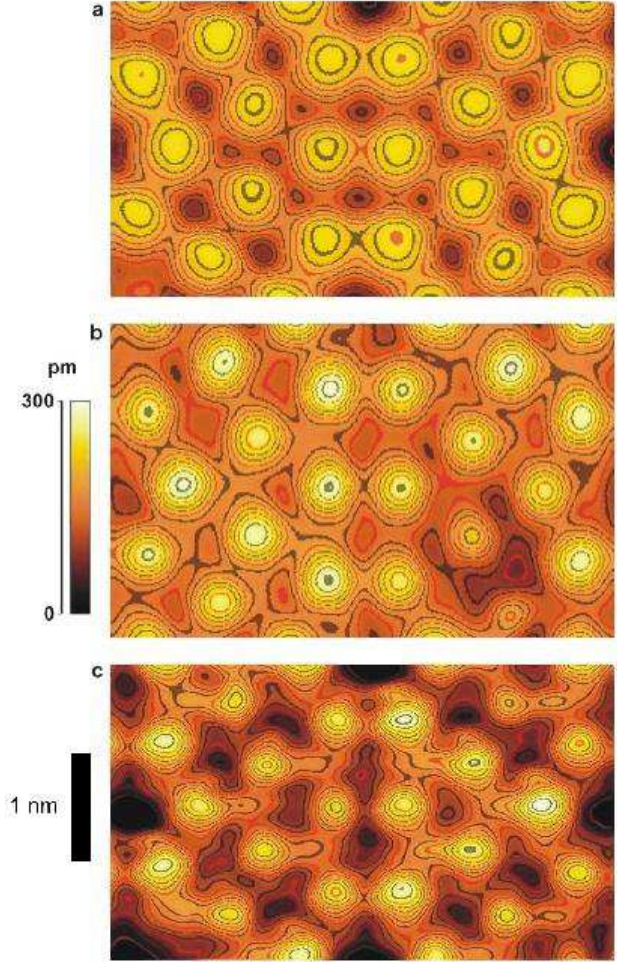}
\caption{\label{fig:Fig5} (Color online) Dynamic STM images of the
Si (111)-(7$\times$7) surface showing the influence of the angular
momentum number $l$ of the tip state to the apex radii. The images
have been taken with one instrument, using similar imaging
parameters but tips made of different materials. The spacing of
the contour lines is 20 pm and their increasing density from (a)
to (c) is caused by decreasing apex radii. (a) Image taken with a
Si tip (a Fe tip suffered a mild collision with a Si sample,
leading to a Fe tip covered with a Si cluster). Sample bias 300
mV, tunneling current $I_{av} =$ 300 pA with $A =$ 0.1 nm. The tip
atom is presumably in a p$_z$-like sp$^3$-state ($l=1$). The
experimental apex radius is approximately 0.5 nm. (b) Image taken
with a Co tip at a sample bias of 200 mV, tunneling current
$I_{av} =$ 100 pA with $A = 0.5$ nm. The image is expected to
originate from a convolution of a Co 3d$_{z^2}$ state ($l=2$) with
the p$_z$ Si states. The apex radius is approximately 0.2 nm. Rest
atoms are visible in the left half of the unit cell. (c) Image
taken with a Sm tip (see also Fig.~\ref{fig:Fig3}c). The rest
atoms are clearly separated from the adatoms. The apex radius is
only 0.14 nm. The tilt angle of the f$_{z^3}$ state is presumably
close to 0$^{\circ}$. Other states (4f, 6s) are possibly
contributing to the image, too.}
\end{figure}

An experimental test of Eqs.~(\ref{eq:rz}-\ref{eq:r7}) can be
performed when imaging a surface with known electronic states with
a tip that is also characterized by a known state. On the Si
(111)-(7$\times$7) surface, the sample states are well
approximated by p$_z$ states. The tip states depend on the
chemical identity of the tip atom. We have therefore looked for
tip materials which exhibit p, d and f states. Si with an
electronic configuration of [Ne]3s$^2$2p$^2$ is a good choice for
p-states, because the 3p states are the highest occupied orbitals
in Si. Co with an electronic configuration of [Ar]3d$^7$4s$^2$ is
expected to have d-symmetry and rare-earth elements are excellent
candidates for f-type tip atoms. We have chosen Sm (electronic
configuration [Xe]4f$^6$6s$^2$) for an f-type tip, because it is
one of the most stable rare-earth elements. For imaging with very
small tip-sample distances, large forces are expected to occur and
mechanical stiffness of the STM tips is important. Rare earth
elements are interesting for STM studies, because in contrast to
hydrogen and other lighter atoms, the 6s electrons are bonded more
strongly to the nucleus than the 4f electrons. Thus, the 4f
electrons participate in electronic conduction and in the
tunneling process from tip to sample. However, the average
distance to the nucleus is larger for the 6s electrons than for
the 4f electrons. Imaging Sm by STM implies that the outer 6s
electron shell has to be penetrated - literally requiring to peek
into the inner parts of an atom.

Fig.~\ref{fig:Fig5} displays the series of dynamic STM images of
p, d and f atoms, realizing the three cases presented in
Fig.~\ref{fig:Fig4} (b)-(d). In Fig.~\ref{fig:Fig5} (a), a Si
front atom is imaged by the Si surface, in Fig.~\ref{fig:Fig5} (b)
a Co atom and in Fig.~\ref{fig:Fig5} (c) a Sm atom. As revealed by
this image, with increasing angular momentum of the tip states,
the apex radii decrease indeed, very much as predicted in
Eqs.~(\ref{eq:rp}-\ref{eq:r7}). Figure~\ref{fig:Fig5} (c) is
recorded with a tip made of pure Sm. The clear visibility of the
rest atoms and the enhanced corrugation together with the very
small apex radius of approximately 0.14 nm prove the angular
confinement resulting from 4f$_{z^3}$ ($l=3$) tip states. Rest
atoms have already been observed by force microscopy
\cite{Lantz,Giessibl2001a,Eguchi}.

\begin{figure}
\includegraphics[width=8.6cm]{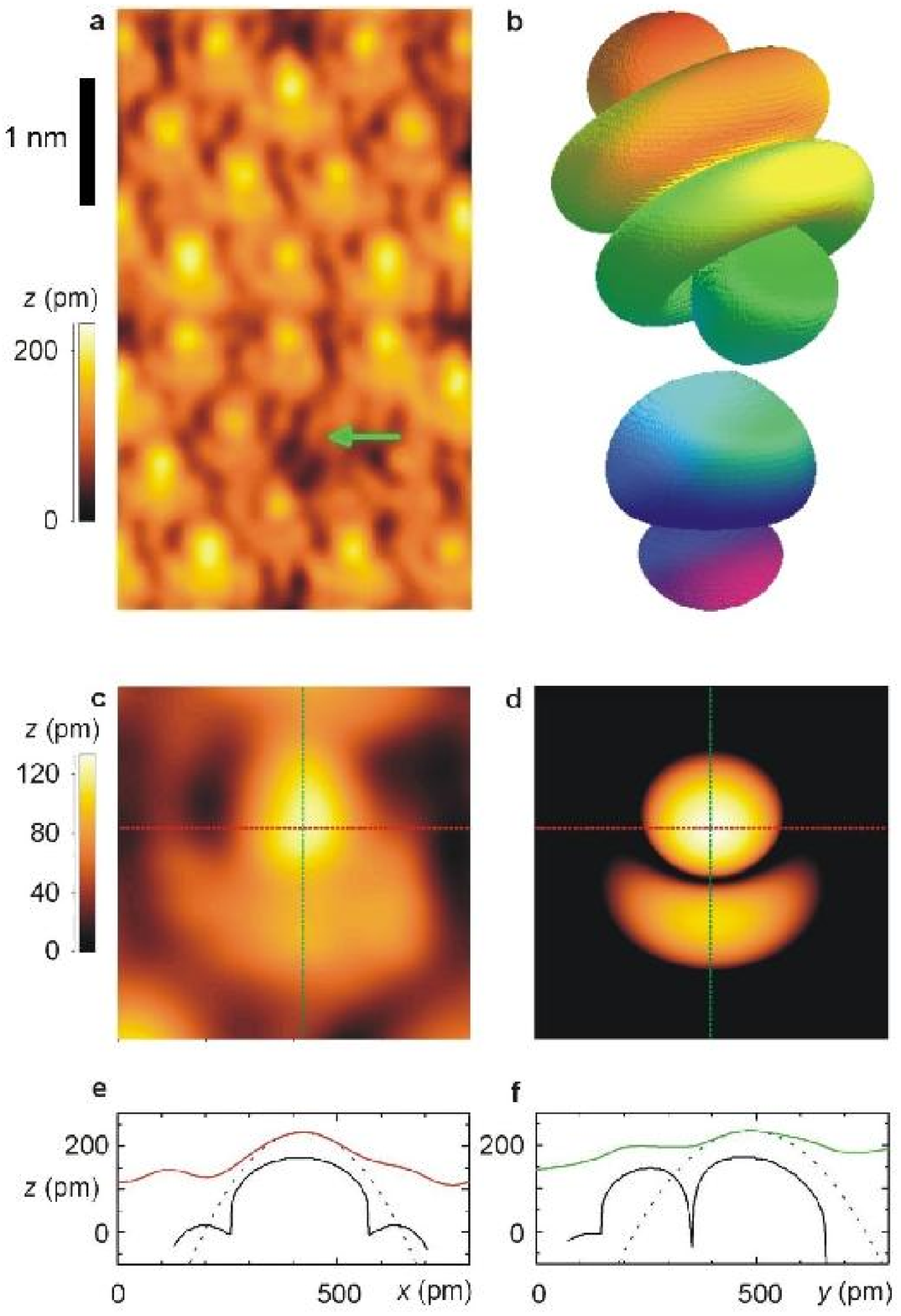}
\caption{\label{fig:Fig6} (Color online) (a) Ultra-high resolution
image of a Si (111)-(7$\times$7) surface, acquired by dynamic STM
with a Co$_6$Fe$_3$Sm tip oscillating with an amplitude of $A =$
0.5 nm at a sample bias voltage of -100 mV and an average
tunneling current $I_{av} =$ 200 pA. The green arrow indicates an
atomic defect (missing center-adatom). (b) Schematic presentation
of the current carrying atomic states leading to the observed
image shown in a. The sp$^3$ silicon states are tunneling mainly
into a Sm 4f$_{z^3}$ tip state tilted by a fixed angle, determined
to be approximately 37$^{\circ}$. (c) Experimental image of a
single Si adatom imaged with a Co$_6$Fe$_3$Sm tip. Average
tunneling current $I_{av}$ = 1 nA, sample bias: -100 mV, amplitude
$A =$ 0.5 nm. (d) Calculated dynamic STM topography image for a
silicon 3p$_z$ sample state and a Sm 4f$_{z^3}$ tip state inclined
37$^{\circ}$ with respect to the $z-$axis. The calculation is
based on the "Modified Bardeen Approach" \cite{Chen} in which the
cantilever oscillation (amplitude $A =$ 0.5 nm) is accounted for.
Average tunneling current $I_{av} =$ 1 nA, sample bias: -100 mV.
The color scales in (c) and (d) are identical. (e) Trace through
maximum of image in $x$-direction (red line). The paraboloid
fitted to the trace has an apex radius of $R_x =$ 0.12 nm. (f)
Trace through maximum of image in $y$-direction (green line). The
paraboloid fitted to the trace has an apex radius of $R_y =$ 0.15
nm. The black lines in (e) and (f) show the corresponding
cross-sections of the constant current surface at the main peak of
the theoretical image. For clarity, the theoretical line-scans are
displaced in $z$-direction.}
\end{figure}

Fig.~\ref{fig:Fig6} (a) shows an experiment where a CoSm
rare-earth magnet mounted on a quartz force sensor was used as a
tunneling tip. The composition of the tip was Co$_6$Fe$_3$Sm, as
determined by Energy Dispersive X-Ray (EDX) analysis. The magnetic
field axis was oriented perpendicular to the sample surface. The
eigenfrequency of the force sensor was $f_0 =$ 19621 Hz, and the
positive frequency shift which was measured during imaging implies
repulsive tip-sample forces during imaging. Qualitatively, the
adatom images in Fig.~\ref{fig:Fig6} (a) can be explained with a
p$_z$ -sample which tunnels into a f- tip state tilted by an angle
$\Theta$ as shown in Fig.~\ref{fig:Fig6} (b). Repulsive forces are
expected to occur when the 6s shell in Sm has to be penetrated in
order to allow for a large tunneling current. A comparison of this
data with images taken with pure Sm, Co and Fe tips suggests that
the tip atom in this image was Sm. The crescent surrounding the
cusp is expected to originate from the upper hoop surrounding the
club of a f$_{z^3}$ orbital as shown in Fig.~\ref{fig:Fig6} (b).
Several investigations were performed to check if these images are
caused by experimental artifacts. Interference due to a multi-tip
image has to be ruled out because of the presence of a single
atomic surface defect indicated by a green arrow in
Fig.~\ref{fig:Fig6} (a) and because of the large depth of the
corner hole image. Further, rotating the fast scanning direction
and varying the scanning speed did not change the images, which
rules out feedback artifacts. The cantilevers frequency shift and
damping were recorded in parallel to the topography record. The
frequency shift according to Fig.~\ref{fig:Fig6} (a) was positive,
i.e. repulsive forces have been acting. The unusually sharp apex
radii can not be explained by atomic relaxations due to tip-sample
forces. While the radius of the atomic images \textit{decreases}
in STM when attractive tip-sample forces act \cite{Hofer}, imaging
at repulsive forces leads to \textit{increasing} radii of the
atomic images, as shown experimentally in Figs. 15 and 16 in
\cite{Giessibl2001a}. In the experiments presented here, repulsive
tip sample forces were acting which cause even an increase in the
experimental apex radii. The damping signal did not show
significant variations within the adatom image, which proves that
a hysteretic lateral jump of the adatom or tip atom which possibly
could have explained the sharp atomic peaks, did not occur
\cite{Giessibl2002}. In theory, fluctuations of the cantilevers
amplitude could explain modulations in topography. However, the
oscillation amplitude was kept constant by an amplitude control
circuit and monitored, ruling out amplitude fluctuations as a
source of the measured atom shapes.

\begin{figure}
\includegraphics[width=8cm]{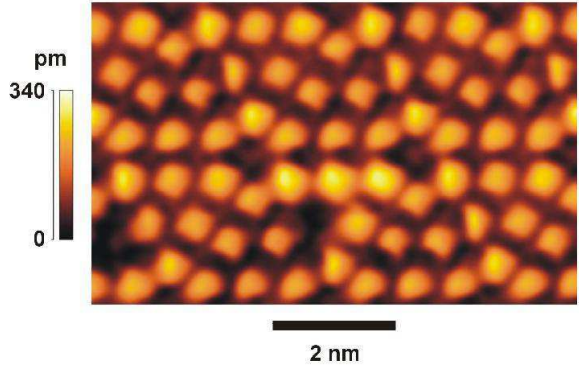}
\caption{\label{fig:Fig7} (Color online) Experimental dynamic STM
image of the Si (111)-(7$\times$7) surface using a samarium tip.
Sample bias -0.8 V, tunneling current $I_{av} =$ 50 pA with $A =$
0.8 nm. Two single atomic defects are present.}
\end{figure}

There is also quantitative theoretical evidence that the observed
features are images of orbitals. Fig.~\ref{fig:Fig6} (c) is a
magnified view of an experimental single adatom image acquired
with the CoSm tip in dynamic STM. Fig.~\ref{fig:Fig6} (d) shows a
calculated image using the "Modified Bardeen Approach"
\cite{Chen}, which is based on quantum mechanical perturbation
theory. For the many-electron-atoms silicon, cobalt, iron and
samarium, "Slater-Type-Orbitals" \cite{McWeeny} were calculated as
approximations for the electronic 3p and 4f states. The only free
input parameter for the calculations is the electronic density of
states at the Fermi level \cite{Chen} of the tip, which was set to
0.4 electrons per atom and eV as a typical value for metals
\cite{Ashcroft1976}. The main result of these calculations is an
excellent agreement between the experiment and theoretical results
for Sm 4f$_{z^3}$ wavefunctions inclined 37$^{\circ}$ with respect
to the axis normal to the silicon surface. Fig.~\ref{fig:Fig6}
(e), (f) show contour lines in $x$ and $y$ direction for both
experimental and calculated images. The calculation accounts for
an oscillation of the cantilever with an amplitude of $A =$ 0.5 nm
in $z-$direction. In experiments as well as in calculations, the
resulting image is insensitive to the amplitude $A$, as long as
$A$ is in the range of 0.3 - 1.5 nm. The difference between the
experimental and theoretical images (a slight compression in
$z-$direction of the experimental image compared to the
theoretical image) is explained with atomic relaxations. Molecular
dynamics calculations reveal that the repulsive interaction
occuring at the lower turning point of the oscillation compresses
the atom image in the $z-$direction, leading to an increased
experimental atom radius. The excellent agreement between
Fig.~\ref{fig:Fig6} (c) and Fig.~\ref{fig:Fig6} (d) implies, that
in this experiment a pure 4f$_{z^3}$ state has carried the major
part of the tunneling current and contributions of other 4f and
possibly 6s states are negligible. Whether the high selectivity
for the 4f$_{z^3}$ state is caused by the magnetic field generated
by the permanent magnet or by crystal field splitting in the field
of the neighbor atoms of the front atom remains to be studied. In
standard STM experiments, much higher tunneling voltages than in
our experiment are usually applied, probably resulting in a
contribution of atomic states of more than one symmetry type (see
Fig.~\ref{fig:Fig3} (a), (b). In this case, it is likely that the
contributions of the various orbitals add to a roughly spherical
symmetry of the image. As another important result of the
calculations, it was found that the oscillation of the tip favors
a monotonic control signal for the $z$-control of the STM
feedback. Therefore an oscillating tip may avoid a tip crash and
enable lateral movement of the probe during the imaging process in
cases, where imaging might be impossible using a static probe. For
static tips, atomic resolution imaging in contact mode should be
degraded because of instabilities of the tip atom. It was
impossible to obtain atomically resolved images with a sample bias
of only -100 mV when reducing the cantilever amplitude below 0.2
nm. Amplitudes considerably larger than the decay length of the
tunneling current result in lower signal, because then electron
tunneling can only occur for a small fraction of the oscillation
cycle. The large spring constant of the cantilever used in our
experiment assures stable operation at amplitudes in the range of
interatomic distances.

In Fig.~\ref{fig:Fig7}, dynamic STM data acquired with a samarium
tip are shown. The special shapes of the atomic images, mainly
with 4-fold symmetry, are supposed to be the consequence of a
convolution of the sp$^3$ silicon states with different samarium
states protruding from a single samarium atom. The special atomic
shapes are repeated in the adjacent unit cells according to the
periodicity of the Si(111)-(7$\times$7) surface. Here, an
interaction with the rest atoms can not be detected. Single atomic
defects in adjacent Si unit cells prove, that a single tip atom
was responsible for the special atomic images.

In summary, we have found a subatomically varying transition
probability for the tunneling process in STM, demonstrating the
capability of dynamic STM to image structures within atoms caused
by atomic orbitals. The observability of the substructures is
attributed to the dynamic STM mode with a cantilever operated at
amplitudes $A$ in the range of 0.3 nm $\lesssim A \lesssim$ 1.5 nm
and a small tunneling bias voltage. Combining dynamic STM with tip
characterization tools like field ion microscopy
\cite{Schirmeisen2000} should further improve our understanding of
the relation between atom shapes and tip states. Theoretical
considerations link the experimental shape of the atoms to the
atomic orbitals participating in the tunneling process. It is
conceivable that chemical identification of the tip atom is
possible with refined calculations of the tunneling current.

\begin{acknowledgments}
We thank S. Hembacher, T. Kopp and C. Laschinger and W. Scherer
for valuable discussions. This work is supported by the
Bundesministerium f\"{u}r Bildung und Forschung (project
EKM13N6918).
\end{acknowledgments}

\end{document}